\documentstyle[prl,aps,twocolumn]{revtex} 
\begin{document} 
\draft
\font\bfmit=cmmib10
\title{Selection Rule in Josephson Coupling between In and $\text {Sr}_2 \text {RuO}_4$}

\twocolumn[
\hsize\textwidth\columnwidth\hsize\csname@twocolumnfalse\endcsname

\author{R. Jin, Y. Liu\cite{byline}} 
\address{Department of Physics, The Pennsylvania State
University, University Park, PA 16802, USA} 
\author{Z. Q. Mao, Y. Maeno}
\address{Department of Physics, Kyoto University, Kyoto 606-8502, Japan}
\address{CREST, Japan Science and Technology Corporation, Kawaguchi, Saitama 332-0012, Japan}

\date{\today} 
\maketitle

\begin{abstract}

We have studied the Josephson coupling between a conventional 
{\it s}-wave superconductor (In) and 
$\text {Sr}_2 \text {RuO}_4$ and found that the coupling 
is allowed in the in-plane direction, but not along the 
{\it c} axis.  This selection rule indicates that 
the symmetry of superconducting order parameter of 
$\text {Sr}_2 \text {RuO}_4$ is either 
{\it p}- or alternatively, purely {\it d}-wave.  
If $\text {Sr}_2 \text {RuO}_4$ is a 
{\it p}-wave superconductor, as strongly favored by other experimental 
evidence, our result suggests that the pairing state of 
$\text {Sr}_2 \text {RuO}_4$ is
$\Gamma^-_5$ with 
{\bf d}({\bf k}) $= {\bf z}({\text k}_{\text x} \pm i{\text k}_{\text y})$, a nodeless state in 
which the spins of the superconducting electrons lie in the 
$\text {RuO}_2$ planes.

\end{abstract}

%suggested PACS numbers 
]

%body, narrow text

\narrowtext

Recently, the first known Cu-free layered perovskite superconductor, 
$\text {Sr}_2 \text {RuO}_4$ \cite{1} ,
 has emerged as a new focus of superconducting materials research.  
The main issue is whether the pairing symmetry of 
$\text {Sr}_2 \text {RuO}_4$ 
is spin triplet with odd parity ({\it p}-wave) 
as predicted theoretically \cite{2}.  A growing body of experimental evidence, including results obtained from muon spin relaxation \cite{3}, NMR 
$1/T_1$ and Knight shift \cite{4}, neutron scattering \cite{5}, 
impurity effect \cite{6}, proximity Josephson 
coupling effect \cite{7}, and specific heat \cite{8} measurements, 
has shown that the pairing symmetry is unconventional, 
most likely {\it p}-wave.  In particular, the flat Knight shift 
observed across the $T_c$ of 
$\text {Sr}_2 \text {RuO}_4$ may be considered as direct evidence for the 
{\it p}-wave pairing state in 
$\text {Sr}_2 \text {RuO}_4$ \cite{4} (see below).  
In addition to the only known {\it p}-wave superconductor, 
$^3 \text {He}$, certain heavy-fermion compounds are also possible candidates for 
{\it p}-wave superconductors.  However, 
$\text {Sr}_2 \text {RuO}_4$ has an advantage that its 
electronic band structure is considerably 
simpler than those of heavy-fermion compounds, making it perhaps a 
more tractable material for demonstrating a {\it p}-wave pairing state.

Assuming a weak spin-orbit coupling, five possible 
{\it p}-wave states are allowed by the crystal symmetry of 
$\text {Sr}_2 \text {RuO}_4$ \cite{2}.  Among them, the 
$\Gamma^-_5$ state, with the 
{\bf d}-vector given by 
{\bf d}({\bf k}) $= {\bf z}({\text k}_{\text x} \pm {\text i}{\text k}_{\text y})$ 
(${\bf z}$ denotes the unit vector along the {\it c} axis), 
is favored by muon spin relaxation \cite{3} and 
NMR Knight shift \cite{4} results.  The physical meaning of 
{\bf d}-vector is as follows.  The magnitude of the 
{\bf d}-vector is the amplitude of 
the superconducting order parameter.  When projected to the direction of the 
{\bf d}-vector, the component of total superconducting electron spins is zero.  
The $\Gamma^-_5$ state for 
$\text {Sr}_2 \text {RuO}_4$ is nodeless with all spins of the 
superconducting electrons lying in the 
$\text {RuO}_2$ planes.  Our experiment on Josephson coupling between an 
{\it s}-wave superconductor, In, and 
$\text {Sr}_2 \text {RuO}_4$ along different crystalline 
orientations, is shown schematically in Fig. \ref{Fig1}. 
This experiment can be used to determine which pairing state within the 
{\it p}-wave scenario is adopted by 
$\text {Sr}_2 \text {RuO}_4$.  In this Letter, we present our experimental 
finding of a selection rule in the Josephson coupling between In and 
$\text {Sr}_2 \text {RuO}_4$. It was found that this coupling is allowed in 
the in-plane direction (Fig. 1a) but not along the {\it c} axis (Fig. 1b).  
If $\text {Sr}_2 \text {RuO}_4$ is a spin-triplet superconductor, 
as strongly favored by other experimental results, our observation provides 
{\it direct} experimental evidence that the pairing state of 
$\text {Sr}_2 \text {RuO}_4$ is indeed 
$\Gamma^-_5$.  In the context of the spin-singlet 
{\it d}-wave scenario \cite{9}, which is not inconsistent with our selection rule but contradicts the NMR Knight shift result \cite{4}, the present work 
suggests that the pairing state in 
$\text {Sr}_2 \text {RuO}_4$ is purely {\it d}-wave.

Single crystals of 
$\text {Sr}_2 \text {RuO}_4$ were grown by a floating-zone method using an image furnace \cite{1}.  Results from {\it a}.{\it c}. magnetic susceptibility 
measurements showed a superconducting transition at 
$T = T_c = 1.45$ K (onset) and a transition width around 0.05 K 
for crystals prepared in two separate growth runs.  
Its superconducting coherence lengths at zero temperature are 
$\xi_{\text {ab}} = 660\,\ \text{\AA}$ and $\xi_{\text c} = 33\,\ \text {\AA}$ for the in-plane and 
{\it c}-axis directions, 
respectively \cite{10}.  To prepare {\it c}-axis 
In/$\text {Sr}_2 \text {RuO}_4$ junctions, 
a $\text {Sr}_2 \text {RuO}_4$ single crystal was cleaved along the 
{\it ab}-plane. Atomic force microscope (AFM) studies of the cleaved surface show an atomically flat surface over an area of up to 
$(10\,\ \mu {\text m})^2$.  A freshly cut In wire of 0.25 mm in diameter was pressed on the crystal immediately after it was cleaved.  The in-plane junctions were prepared on 
$\text {Sr}_2 \text {RuO}_4$ single crystals with a finely polished {\it ac} face. AFM imaging showed that the polished face is fairly rough with micron-size mechanical damage.  To our knowledge, no chemical solution can etch 
$\text {Sr}_2 \text {RuO}_4$ to obtain a smooth surface.  
Thus, the freshly cut In wire was directly pressed on clean, but as-polished {\it ac} face of $\text {Sr}_2 \text {RuO}_4$ to form an in-plane junction.  
For both types of junctions, the junction area is 
$\sim \text {0.05 mm}^2$ with junction resistances ranging from 0.1 to 
100 $\Omega$.  Electrical measurements were carried out in {\it d}.{\it c}. in a 
$^3 \text {He}$ cryostat with a base temperature of 0.3 K.  
A $\mu$-metal box shielded the samples from residual magnetic field.

In Fig. \ref{Fig2}, the I-V curves of an in-plane 
In/$\text {Sr}_2 \text {RuO}_4$ junction (Sample \#11) are shown. 
In this case, the current ($I$) flows along the in-plane direction. A non-zero supercurrent, followed by a linear I-V characteristic, was evident.  Qualitatively the same behaviors have been found in four other in-plane junctions with non-zero critical current 
$I_{\text c}$.  The temperature dependence of 
$I_{\text c}$, shown in 
Fig. \ref{Fig3} for Samples \#11 and \#12, has the general shape of that for a superconductor-normal metal-superconductor (SNS') junction \cite{11}. 
The magnetic field dependence of 
$I_{\text c}$ was measured for one in-plane junction (without 
$\mu$-metal shield).  While 
$I_{\text c}$ was found to decrease with increasing field, no Fraunhofer pattern was observed, suggesting that the junction is not very uniform.  For two dissimilar 
{\it s}-wave superconductors, the Ambegaokar-Baratoff (A-B) limit for 
$I_{\text c}R_{\text N}$ is given by 
$I_{\text c}R_{\text N} \leq (\Delta_1/e) {\text K}\{[1-(\Delta_1/\Delta_2)^2]^{1/2}\}$, 
where $R_{\text N}$ is junction resistance in the normal state,
$\Delta_1$ and $\Delta_2$ are zero-temperature energy gaps for two superconductors, 
and the function K is the elliptic integral of the first kind \cite{12}.  Unfortunately, the gap for $\text {Sr}_2 \text {RuO}_4$ is yet to be determined 
experimentally \cite{13}.  However, if one estimates the gap using the BCS result, 
$\Delta = 1.764{\text k}_{\text B}T_{\text c} \approx$ 0.22 meV, 
this leads to an A-B limit of 0.6 mV.  
At $T = 0.3$ K, values of 
$I_{\text c}R_{\text N}$ are 0.16 and 0.18 mV for Junctions \#11 and \#12, 
and 15, 16, and 45 $\mu$V for three other (in-plane) samples, respectively.  The numbers for Junctions \#11 and \#12 are a substantial fraction of the A-B limit, suggesting that, at least for these two junctions, the observed $I_c$ is due to a finite Josephson coupling between In and $\text {Sr}_2 \text {RuO}_4$ in the in-plane direction.  

In {\it c}-axis In/$\text {Sr}_2 \text {RuO}_4$ junctions, 
no finite supercurrent was found.  While it is not clear why a natural barrier is easier to form in {\it c}-axis junctions, experimentally most 
{\it c}-axis junctions were found to exhibit tunneling behavior.  
In principle, it is possible that the presence of a tunnel barrier suppresses the amplitude of the superconducting order parameter \cite{14}, resulting in a vanishing 
supercurrent, independent of the pairing symmetry of 
$\text {Sr}_2 \text {RuO}_4$.  In two {\it c}-axis 
In/$\text {Sr}_2 \text {RuO}_4$ junctions, however, 
instead of tunneling features, an excess current 
($I_0$) or zero-bias conductance peak (Fig. \ref{Fig4}) was seen.  The excess current or zero-bias conductance peak is a signature of the Andreev reflection process at a normal metal-superconductor (N-S) interface, where an incoming normal electron with energy below the gap of the superconductor combines with another electron to form a Cooper pair which enters the superconductor.  As a result, a hole is reflected, giving rise to "extra" charge passing through the interface.  Since the Andreev reflection occurs only at a metallic interface, its presence above the 
$T_{\text c}$ of $\text {Sr}_2 \text {RuO}_4$ indicates that the interface between In and $\text {Sr}_2 \text {RuO}_4$ is metallic in nature.  
This excess current was seen to persist below the 
$T_{\text c}$ of $\text {Sr}_2 \text {RuO}_4$, which may be explained by 
the existence of a normal layer at the {\it ab} face of the 
$\text {Sr}_2 \text {RuO}_4$ crystal.  This is further supported 
by the observation that no gap features associated with 
$\text {Sr}_2 \text {RuO}_4$ were present below the 
$T_{\text c}$ of $\text {Sr}_2 \text {RuO}_4$ in all {\it c}-axis 
tunnel junctions we have studied \cite{13}.  Although the precise origin of this normal layer is unknown, it may be due to mechanical stress and/or oxygen deficiency near the crystal surface.  No supercurrent was observed down to 0.38 and 
0.65 K respectively in either of these two 
{\it c}-axis junctions showing Andreev reflection, suggesting that no 
Josephson coupling was established in these {\it c}-axis junctions.

An important question is whether the lack of finite Josephson coupling between In and 
$\text {Sr}_2 \text {RuO}_4$ along the {\it c} axis is of an intrinsic or 
extrinsic origin.  In addition to a metallic contact between In and the normal top layer of $\text {Sr}_2 \text {RuO}_4$, the interface between the 
normal layer and the bulk superconducting 
$\text {Sr}_2 \text {RuO}_4$ should be metallic as well since 
it is naturally formed with no oxygen deficiency 
or mechanical stress expected.  As a result, the two 
{\it c}-axis In/$\text {Sr}_2 \text {RuO}_4$ junctions 
showing Andreev reflection should be SNS' junctions with two metallic interfaces.  If 
$\text {Sr}_2 \text {RuO}_4$ is an {\it s}-wave superconductor, 
Josephson coupling should be possible in these 
{\it c}-axis junctions as long as the thickness of the N-layer is within a few times of the normal coherence length $\xi_{\text N}$ \cite{11}.  In the clean limit, 
which we believe is appropriate for 
$\text {Sr}_2 \text {RuO}_4$, 
$\xi_{\text N} = \hbar v_{\text N}/(2 \pi {\text k}_{\text B}T)$, where 
$v_{\text N}$ is the Fermi velocity.  Using $v_{\text N} = 1.4 \times 10^6$ cm/s along the 
{\it c} axis \cite{15}, we have 
$\xi_{\text N} =$ 774 ${\text \AA}$ for 
$\text {Sr}_2 \text {RuO}_4$ at $T = 0.38$ K, the lowest temperature 
measured for Sample \#17.  Compared with the distance between two adjacent 
$\text {RuO}_2$ layers, 6.4 \AA, a length a few times of 
$\xi_{\text N}$ would be of a few hundreds of the inter-layer distance.  It is very unlikely that the normal layer formed at a freshly cleaved 
$\text {Sr}_2 \text {RuO}_4$ single crystal can be so thick.  
Therefore the absence of supercurrent in 
{\it c}-axis In/$\text {Sr}_2 \text {RuO}_4$ junctions cannot be 
due to an overly thick N-layer.

Is it possible that the supercurrent is too small to be measured for these 
{\it c}-axis junctions?  If the A-B limit for 
$I_{\text c}R_{\text N}$ (0.6 mV) is a good guide, we expect a critical current on the order of 100 $\mu$A for Sample \#17 with 
$R_{\text N}$ = 5.7 $\Omega$.  Even using the experimental values of 
$I_{\text c}R_{\text N}$ for two in-plane 
In/$\text {Sr}_2 \text {RuO}_4$ junctions shown in Fig. \ref{Fig3} (0.16 and 0.18mV respectively at 0.3 K), we still expect 
$I_{\text c} >$ 28 $\mu$A, well above our measurement limit.  For another 
{\it c}-axis junction which also showed Andreev reflection features, 
$R_{\text N} =$ 0.22 $\Omega$, we should expect an even larger supercurrent.  
Therefore, the absence of Josephson coupling between In and 
$\text {Sr}_2 \text {RuO}_4$ along 
{\it c} axis must be due to intrinsic reasons, probably a spin-triplet pairing state in 
$\text {Sr}_2 \text {RuO}_4$.  It should be noted that, unlike experiments on 
{\it c}-axis tunneling between Pb and high-$T_{\text c}$ 
superconductors \cite{16,17}, we are attempting to demonstrate 
the absence, not the presence, of a 
{\it c}-axis Josephson coupling between the two superconductors.  
Hence, whether or not the in-plane tunneling due to the presence of steps on the cleaved {\it ab} face may be present in our 
{\it c}-axis junctions is not an issue since the presence of the 
in-plane coupling will only help give rise to a finite $I_{\text c}$.

The absence of Josephson coupling between a spin-singlet 
(even-parity) and a spin-triplet (odd-parity) superconductor was first proposed as a test for the unconventional pairing state in heavy-fermion superconductors \cite{18}.  
However, it was subsequently pointed out that the first order Josephson coupling between two superconductors with different parities was possible through spin-orbit 
coupling \cite{19}.  In the presence of the spin-orbit coupling, 
the Cooper pairs of different parities will be mixed at the interface between the 
{\it s}- and the {\it p}-wave superconductor, 
resulting in a direct Josephson coupling between them.  
In fact, it has been shown that $I_{\text c}$ can be written as \cite{20}
	\begin{equation}
$$I_{\text c} = 2e \, \text {Im} \int \lambda T_{\text t}^2 F^{\ast} K
	({\bf x}) [({\bf d}({\bf k},{\bf x}) \times {\bf k})
	$$\cdot {\bf n}]$${\text d}{\bf x}{\text d}{\bf k}$$
	\eqnum{1}
	\end{equation}
where $\lambda$ is a dimensionless parameter, 
$T_{\text t}$ the tunneling matrix, 
$F^{\ast}$ the Gor'kov function, $K$ the kernel and {\bf n} the unit vector normal to the interface.  
In Eq. 1, {\bf x} and {\bf k} are real- and momentum-space coordinates, 
and the integration is carried out over the junction interface and the Fermi surface of the {\it p}-wave superconductor.  The physical origin of Eq. 1 can be easily understood if we rewrite  
[({\bf d}({\bf k},{\bf x})$\times${\bf k})$\cdot {\bf n}]$ as 
[{\bf d}({\bf k},{\bf x})$\cdot$({\bf k}$\times {\bf n})]$.  Since 
{\bf d} and {\bf k} $\times {\bf n}$ represent essentially the spin 
and the orbital angular momentum of the superconducting 
condensate at the interface, respectively,  
[{\bf d}({\bf k},{\bf x})$\cdot$({\bf k}$\times {\bf n})]$ 
merely reflects the spin-orbit coupling strength of the 
{\it p}-wave superconductor, which gives rise to the (orientation-dependent) Josephson coupling between the 
{\it p}- and the {\it s}-wave superconductor as mentioned above.  
If the pairing symmetry in 
$\text {Sr}_2 \text {RuO}_4$ is indeed 
{\it p}-wave, Eq. 1 implies that the Josephson coupling between In and 
$\text {Sr}_2 \text {RuO}_4$ is orientation-dependent.  
In particular, among five possible representations 
($\Gamma^-_{1-5}$) \cite{2}, Eq. 1 states that, for in-plane junctions, 
$I_{\text c} \neq 0$ only if the pairing state of 
$\text {Sr}_2 \text {RuO}_4$ is 
$\Gamma^-_5$ with 
{\bf d}({\bf k}) $= {\bf z}({\text k}_{\text x} \pm i{\text k}_{\text y})$.

The above result is consistent with {\it all other} experimental findings obtained thus far.  In particular, in the NMR experiment \cite{4}, the electron spin susceptibility of 
$\text {Sr}_2 \text {RuO}_4$ was found to be a constant 
within experimental error as temperature was brought from above 
$T_{\text c}$ to 15 mK, as expected for the
$\Gamma^-_5$ state.  While the standard theory predicts 
exponentially small electron spin susceptibility in the zero-temperature limit for 
{\it s}-wave superconductors \cite{21}, it was found that Hg \cite{22} and 
Sn \cite{23} showed finite electron spin susceptibility well below 
$T_{\text c}$.  This was explained \cite{24} by the presence of 
spin-orbit coupling within an {\it s}-wave picture.  
Nevertheless, a finite drop in electron spin susceptibility was still observed across 
$T_{\text c}$ for both Hg \cite{22} and Sn \cite{23}.  
A constant electron spin susceptibility (or Knight shift) in 
$\text {Sr}_2 \text {RuO}_4$ containing relatively light 
elements (corresponding to weak spin-orbit coupling) is 
difficult to be explained within a spin-singlet picture.

It should be pointed out that a {\it d}-wave scenario has recently been proposed for 
$\text {Sr}_2\text {RuO}_4$ \cite{9}.  This scenario contradicts the NMR Knight shift result.  However, if it turns out to be true, the result of 
the present work suggests that the superconducting order parameter in 
$\text {Sr}_2 \text {RuO}_4$ is purely {\it d}-wave. 
Given that the issue of whether the pairing symmetry in high-
$T_{\text c}$ cuprates is purely {\it d}-wave is still being debated, 
our selection rule result would be of significance in that context as well.

In conclusion, we have found a selection rule in the Josephson coupling between In and 
$\text {Sr}_2 \text {RuO}_4$.  While a phase sensitive experiment may ultimately be required to settle this issue, the totality of the available results on 
$\text {Sr}_2 \text {RuO}_4$, and the remarkable consistency among them, 
makes a compelling case for a {\it p}-wave pairing state in 
$\text {Sr}_2 \text {RuO}_4$.  In this context, the present 
work shows that the pairing state of 
$\text {Sr}_2 \text {RuO}_4$ is 
$\Gamma^-_5$.

We would like to acknowledge helpful discussions with D.F. Agterberg, A.J. Millis, T-L. Ho, J. Sauls, and C.C. Tsuei, and particularly M. Sigrist who has brought our attention to Eq. 1. This work is supported in the US In part by NSF through grants DMR-9702661, DMR-9974327, and ECS-9705839.

%references

%figures

\begin{figure} 
\caption{Schematics of (a) in-plane  and (b) {\it c}-axis  
In/$\text {Sr}_2 \text {RuO}_4$ junctions.} 
\label{Fig1}
\end{figure}

\begin{figure} 
\caption{I-V curves at various temperatures for an In-plane 
In/$\text {Sr}_2 \text {RuO}_4$ junction (Sample \#11).  
Finite critical current $I_{\text c}$ is indicated.}
\label{Fig2} 
\end{figure}

\begin{figure} 
\caption{Temperature dependence of the critical current, 
$I_{\text c}(T)$, for two in-plane 
In/$\text {Sr}_2 \text {RuO}_4$ junctions (samples \#11 and \#12).}  
\label{Fig3} 
\end{figure}

\begin{figure} 
\caption{a) I-V curves for a {\it c}-axis 
In/$\text {Sr}_2 \text {RuO}_4$ Junction (Sample \#17); 
b) Dynamic conductance of the same junction.}
\label{Fig4} 
\end{figure}


\begin{references}
\bibitem[*]{byline} To whom correspondence should be addressed.
\end{references}
\end{document}